\documentclass[9pt,conference]{IEEEtran}
\IEEEoverridecommandlockouts
\usepackage{cite}
\usepackage{amsmath,amssymb,amsfonts}
\usepackage{graphicx}
\usepackage{bm}
\usepackage{booktabs} % \toprule \midrule \bottomrule \cmidrule{start-end} 表格的 start 到 end 列添加粗线
\usepackage{diagbox} % 创建具有斜线表头的表格单元格 \diagbox{Col 1}{Col 2}.
\usepackage{enumitem}
\usepackage{multicol}
\usepackage{multirow}
\usepackage{makecell}
\usepackage{pifont}
\usepackage{xcolor}
\usepackage{graphicx}
\usepackage{stfloats}
\usepackage{booktabs}
\usepackage{url}
\usepackage{dashrule}
\usepackage[linesnumbered,lined,ruled,commentsnumbered]{algorithm2e}
\usepackage{url,hyperref,cleveref}

\def\BibTeX{{\rm B\kern-.05em{\sc i\kern-.025em b}\kern-.08em
    T\kern-.1667em\lower.7ex\hbox{E}\kern-.125emX}}

\begin{document}

\title{Neural Directed Speech Enhancement \\ with Dual Microphone Array in High Noise Scenario}

\author{\IEEEauthorblockN{Wen Wen$^1$, Qiang Zhou$^{1,2}$, Yu Xi$^1$, Haoyu Li$^1$, Ziqi Gong$^2$, Kai Yu$^{1{\dagger}}$ \thanks{$^{\dagger}$Kai Yu is the corresponding author.} }
\IEEEauthorblockA{\textit{$^1$MoE Key Lab of Artificial Intelligence, AI Institute, X-LANCE Lab, Shanghai Jiao Tong University, Shanghai, China}}
\IEEEauthorblockA{\textit{$^2$AISpeech Ltd, Suzhou, China}}
\IEEEauthorblockA{
\{wenyideng, yuxi.cs, haoyu.li.cs, kai.yu\}@sjtu.edu.cn~~~ \{qiang.zhou, ziqi.gong\}@aispeech.com
}

}
\maketitle

\begin{abstract}

In multi-speaker scenarios, leveraging spatial features is essential for enhancing target speech. While with limited microphone arrays, developing a compact multi-channel speech enhancement system remains challenging, especially in extremely low signal-to-noise ratio (SNR) conditions. To tackle this issue, we propose a triple-steering spatial selection method, a flexible framework that uses three steering vectors to guide enhancement and determine the enhancement range. Specifically, we introduce a causal-directed U-Net (CDUNet) model, which takes raw multi-channel speech and the desired enhancement width as inputs. This enables dynamic adjustment of steering vectors based on the target direction and fine-tuning of the enhancement region according to the angular separation between the target and interference signals. Our model with only a dual microphone array, excels in both speech quality and downstream task performance. It operates in real-time with minimal parameters, making it ideal for low-latency, on-device streaming applications.

\end{abstract}


\section{Introduction}
Recently, speech enhancement~(SE) in extremely low signal-to-noise ratio (SNR) environments has gained significant attention due to its critical applications in telecommunications, hearing aids, keyword spotting~(KWS), and automatic speech recognition~(ASR) systems~\cite{back1,back2,yuxi_2022text,yuxi4,back3,yuxi3,yuxi5,li24r_interspeech,back4}. Although some downstream works~\cite{yuxi1,yuxi2} aim to achieve strong noise robustness, it's difficult to process all kinds of complex challenges without SE front-ends. The main challenges include background noise, interference, and reverberation, all of which can severely degrade speech intelligibility and quality. Among these, human speech interference presents the greatest challenge, as downstream systems struggle to differentiate target speech from competing speech, resulting in a significant performance decline.

It is well known that humans can focus on specific directions using binaural spatial information. Similarly, modern hearing aids, speech recognition systems, and other devices are equipped with multiple microphones. As a result, multichannel SE has gained importance~\cite{importance}, with many researchers exploring the use of spatial features for speech separation or speaker extraction~\cite{explore1,multi_pass,explore3,explore4,explore5}. Traditional multichannel SE techniques, such as beamformers~\cite{bf1,bf2,bf3} like the delay-and-sum~\cite{delay_sum} and generalized sidelobe canceller (GSC)~\cite{GSC}, face performance limitations. Recently, many neural network methods have outperformed traditional methods in both quality and intelligibility of enhancing speech, such as EMGSE~\cite{EMGSE}, VSEGAN~\cite{VSEGAN}, METRICGAN-U~\cite{metricgan}, HGCN~\cite{hgcn}, TF-GridNet~\cite{TF-GridNet} and FullSubNet+~\cite{fullsubnet}. However, they also have several limitations: Firstly, they often predefined the target speech region rather than input spatial information while using, such as GSENet~\cite{gsenet}, BASNet~\cite{BASNet} and DSENet~\cite{dsenet}. Secondly, they are mainly focused on scenarios involving more than three microphones like Multi-pass extraction~\cite{multi_pass}, JNF~\cite{JNF} and JNF-SSF~\cite{jnf_ssf}. Using too many microphones is impractical for on-device SE systems, as resource consumption increases rapidly with the number of microphones, while these systems are constrained by memory and computational resources. Additionally, the large-scale parameters of neural networks limit their feasibility in real-world applications. Lastly, previous works have primarily focused on improving speech quality metrics, often overlooking their impact on downstream tasks.

In this paper, we propose a causal-directed U-Net (CDUNet) model that integrates U-Net~\cite{unet} with beamforming. Our approach focuses on using the spatial location of the target speaker as a cue, aiming to create a non-linear filter that can be flexibly steered in the chosen direction. Additionally, we introduce variable enhancement widths for different interference angles, allowing the model to extract detailed information about interfering signals and dynamically adjust the enhancement range. 

The contributions of this work are summarized as follows:

\begin{enumerate}
\vspace{-1pt}
    \item We propose a novel triple-steering spatial selection method that integrates three steering vectors with a U-Net architecture to determine both the target direction and enhancement scope.
    \item This is the first work to introduce width as an input parameter, enabling the model to flexibly adapt the enhancement range based on the application context.
    \item Compared to conventional beamformers and various neural network baselines, the proposed CDUNet not only delivers superior front-end performance in both fixed and variable target directions but also enhances backend ASR performance.
\end{enumerate}

\section{Triple-steering spatial selection method}

\subsection{Problem Definition}
% \vspace{-3pt}
This research targets the so-called cocktail-party problem: extracting the speech signal of a target speaker from the interfering speech.
We assume that the interfering speakers originate from different directions than the target speaker. Thus, the problem can be described as follows:
\begin{equation}
\vspace{-1pt}
y_i(t) = s(t;\mu_1)+i(t;\mu_2)+n(t),
\end{equation}
where $y_i(t)$ denotes the input signal captured by the $i_{th}$ microphone, $s(t)$ represents the target speech signal, $i(t)$ indicates the interfering speech, $n(t)$ stands for the noise, and $\mu_1$ and $\mu_2$ represent the azimuthal angles of the target and interfering sound sources relative to the microphone, respectively. The goal of speech enhancement is to train a deep neural network $f_{\theta}$ that maps $y(t)$ to $s(t)$. The enhanced signal $\hat{s}$ is then estimated by:
% okk，先这么写，符号后面再说
% 我感觉是吧
% \vec{\bm{y}} = [y_1,y_2,\cdots, y_n]
\begin{equation}
\hat{s} = f(y_{1\sim 2},\mu_1;\theta),
\end{equation}
where $\theta$ represents the network parameters, $\hat{s}$ denotes the predicted output, and $y_{1\sim 2}$ refers to the 2-channel speech signals captured by the microphones.
 
\subsection{Triple-steering Spatial Selection}
\vspace{-3pt}
% In our method, we utilize the target angle and two edge angles that are determined by subtracting and adding the input width to the target angle to generate three steering vectors for the beamformer. Then, the network takes the frequency-domain representations of two raw microphone signals, along with the beamformer outputs at both the target angle and the two edge angles as input. This enables the model to accurately localize the target speaker's direction. Furthermore, by incorporating the input width, the model estimates the angular separation between the target and interference sources, allowing for more precise directed enhancement.

In our method, we utilize the target angle and two edge angles, calculated by subtracting and adding the input width to the target angle, to generate three steering vectors for the beamformer. The network then takes the frequency-domain representations of two raw microphone signals, along with the beamformer outputs at the target angle and the two edge angles as input. This allows the model to accurately localize the target speaker's direction. Additionally, by incorporating the input width, the model estimates the angular separation between the target and interfering sources, enabling more precise directed enhancement.

\vspace{-2pt}

\subsection{Causal-Directed U-Net Model}
\vspace{-3pt}
\begin{figure*}[th]
    \centering
    % \vspace{-45pt}
    \includegraphics[width=1.9\columnwidth]{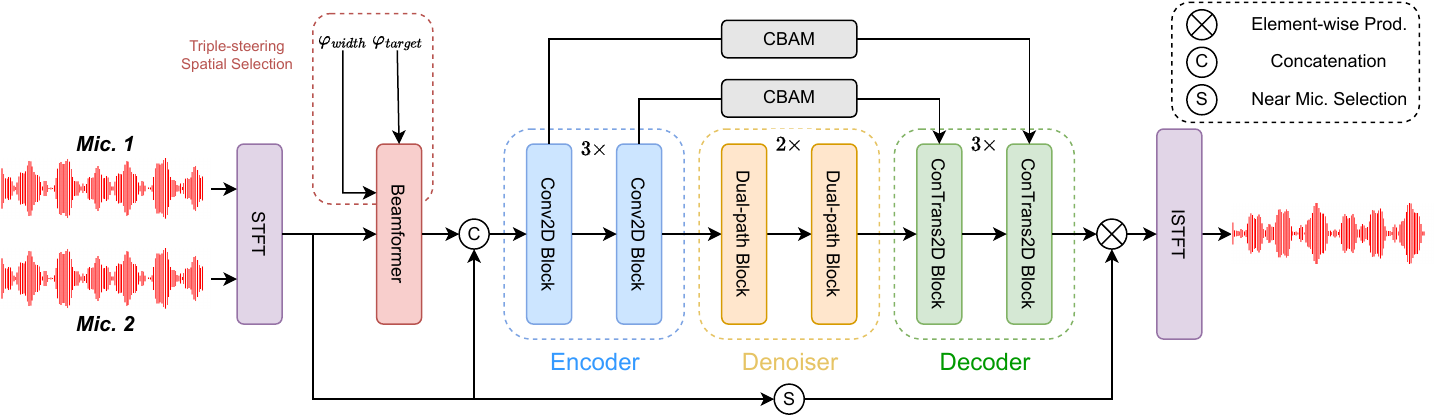}
    % \vspace{-43pt}
    \caption{\textbf{Illustration of the CDUNet architecture.} The beamformer output incorporates both the target direction and the width input, which captures the spatial area information crucial for enhancement. $\varphi_{width}$ denotes the extent of the target region to be enhanced, and $\varphi_{target}$ specifies the orientation of the target speaker. The "Near Mic. Selection" operation selects the speech signal from the microphone that is positioned closer to the target speaker.} 
    \label{overwiew}
\vspace{-1.0em}
\end{figure*}
As illustrated in Figure \ref{overwiew}, our model utilizes a convolution U-Net architecture. We employ a robust encoder-decoder architecture with skip connections to construct a deep neural network. The encoder comprises 3 two-dimensional convolutional (Conv2D) blocks that encode the input features into a latent representation. Correspondingly, the decoder composes 3 two-dimensional transposed convolutional (ConTrans2D) blocks that decode the latent space features back into the feature space.

Between the encoder and decoder, we integrate sequence modeling modules, including a frequency sequence layer and a Long Short-Term Memory(LSTM) layer, following the Dual-Path Recurrent Neural Network(DPRNN) framework. 

Additionally, we incorporate the Convolutional Block Attention Module(CBAM)~\cite{cbam}, which combines channel and spatial attention mechanisms. In our model, CBAM is applied within the decoder and skip connections to recalibrate the Time-Frequency (TF) feature maps, thereby enhancing target reconstruction accuracy.

Let $Y^{(i)} \in \mathbb{R}^{B \times C_i \times F_i \times T}$ be the output of the $i$-th encoder block, where $i \in [1, 3]$. The channel attention gate for the $i$-th block, $G(i)_c \in \mathbb{R}^{B \times C_i \times 1 \times 1}$, is derived as follows:
\begin{equation}
G\left( i \right) _c=\sigma \left( W_2\left( W_1\left( F\left( i \right) _{avg_c} \right) \right) +W_2\left( W_1F\left( i \right) _{\max_c} \right) \right),
\end{equation}
where $\sigma$ represents the sigmoid activation function, while $W_1$ and $W_2$ denote the weights of shared linear layers. The average pooling $F(i){avg_c}=\text{AvgPool}(Y^{(i)})$ and max pooling $F(i){max_c}=\text{MaxPool}(Y^{(i)})$ are applied to $Y^{(i)}$. and the channel attention gate is applied to $Y(i)$ to calculate $Y(i)_c = G(i)_c\cdot Y(i)$:
The spatial attention gate $G_f^{(i)} \in \mathbb{R}^{B \times 1 \times F_i \times T}$ is calculated as follows:
\begin{equation}
G\left( i \right) _f=\sigma \left( fconv\left( \left[ F\left( i \right) _{avgf};F\left( i \right) _{max f} \right] \right) \right),
\end{equation}
where $f_{\text{conv}}$ is a convolutional(Conv2D) layer. The spatial attention gate is then applied to $Y(i)_c$ to calculate $Y_f(i) = G(i)_f \cdot Y(i)_c$.

The output speech signal is estimated by applying the mask generated by the decoder to the nearer raw channel. The nearer channel is selected based on its proximity to the input target angle, with the closer channel chosen for processing. For instance, if the input angle is smaller than 90$^{\circ}$, the first channel is selected as the nearer channel, otherwise, the second channel is used.

\vspace{-5pt}
\subsection{Loss Function}
\vspace{-3pt}
In the Short-Time Fourier Transform(STFT) domain, the target and estimated outputs are denoted as $\mathbf{s}$ and $\hat{\mathbf{s}}$, respectively. Our preliminary studies indicated that employing the Scale-Invariant Signal-to-Noise Ratio(SI-SNR)~\cite{sisnr} loss function significantly enhances the stability of the learning process. However, when the network was trained solely with the SI-SNR loss, we observed a tendency for the network to excessively suppress low-frequency components. To mitigate this issue, we propose an innovative combined loss function incorporating both SI-SNR and Multi-Resolution STFT(MR-STFT)~\cite{MR_STFT} losses. 
The formulation of our proposed combined loss function is as follows:
\begin{equation}
% \mathcal{L} \left( \mathbf{s},\hat{\mathbf{s}} \right) =\alpha _1\sum_{i\in I}{\mathcal{L}_{MR-STFT}^{(i)}}+\alpha _2\mathcal{L}_{SI-SNR}(\mathbf{s},\hat{\mathbf{s}}),\label{3-6}
\mathcal{L} \left( \mathbf{s},\hat{\mathbf{s}} \right) =\alpha _1\sum_{i\in I}{\mathcal{L}_{MR-STFT}^{(i)}}+\alpha _2\mathcal{L}_{SI-SNR}(\mathbf{s},\hat{\mathbf{s}}),\label{3-6}
\end{equation}
where $I$ is the set of STFT points, $\alpha_1$ and $\alpha_2$ are weighting factors.
\begin{figure}[b]
    \vspace{-18pt}
    \hspace{-22pt}
    \begin{minipage}[t]{0.75\columnwidth} % 左侧的图片占据半列
    \centering
    %\vspace{-12pt}
    \includegraphics[width=0.8\linewidth]{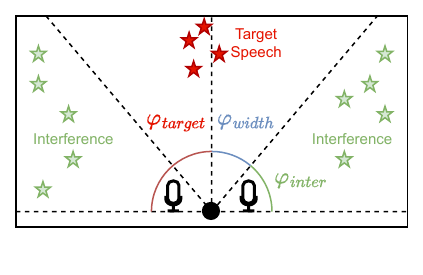}
    % \vspace{-15pt}
    % \caption*{Illustration of Data Simulation}
    \end{minipage}
    % \hfill % 添加水平间距，使两个 minipage 紧靠
    \hspace{-12pt}
    \begin{minipage}[b]{0.2\columnwidth} % 右侧的表格占据半列
    \centering
        \begin{tabular}{lc}
            \toprule
             \multicolumn{2}{c}{Room characteristics} \\
            \midrule
            Width & 2.5-5m \\
            Length & 3-9m \\
            Height & 2.2-3.5m \\
            T60 & 0.2-0.5s \\
            \bottomrule
        \end{tabular}
        \vspace{23pt}
        % \caption*{Room info.}
    \end{minipage}
    \vspace{-17pt}
    \caption{Illustration of the simulation setup of the first fixed-target dataset. The target direction ranges from 85° to 95°, represented by red stars in the figure, while the interference direction is located 15° away from the target direction, indicated by green stars. Room information is uniformly sampled from the provided ranges in the table.}
    \label{simulator}
\end{figure}

\section{Experimental Setup}
\subsection{Dataset}

All of the clean and interference utterances are randomly sampled from the LibriSpeech~\cite{LibriSpeech} and our internal corpora. We simulate training datasets in the following style:

\begin{itemize}
    \item \textbf{Geometry.} Rooms with different geometries are generated in alignment with the JNF~\cite{JNF} benchmark, as depicted in figure \ref{simulator}. Within each simulated room, the microphone spacing is set to 30 mm. The audio sources are selected at random, with one designated as the target source and the other as the interference.
    %Rooms with different geometries are generated following predefined distributions. In alignment with the JNF~\cite{JNF} benchmark, we configured the room characteristics to be identical to those used in JNF as depicted in figure \ref{simulator}. Within each simulated room, the microphone spacing is set to 30mm. The audio sources are selected at random, with one designated as the target source and the other as the interfering source.
    \vspace{-5pt}
    \begin{table*}[t]
    \centering
    \caption{\textbf{PESQ scores for a fixed area} (training filter for target angle around $90^{\circ}$). $\varphi_{inter}$ indicates the direction of interference speech, and Noisy Sp. denotes noisy speech.}       
    \begin{tabular}{c|c|c|ccccccccccccc|c}
        \toprule
        \multirow{2}{*}{\textbf{SNR}} & \multirow{2}{*}{\textbf{UNet}} & \multirow{2}{*}{\textbf{Model}} & \multicolumn{13}{c|}{$\bm{\varphi_{inter}}$} & \multirow{2}{*}{\textbf{Avg.}} \\
        \cmidrule{4-16}
         & & & $0^{\circ}$ & $15^{\circ}$ & $30^{\circ}$ & $45^{\circ}$ & $60^{\circ}$ & $75^{\circ}$ & $90^{\circ}$ & $105^{\circ}$ & $120^{\circ}$ & $135^{\circ}$ & $150^{\circ}$ & $165^{\circ}$ & $180^{\circ}$ & \\
        \midrule
        \multirow{9}{*}{0 dB}& - & Noisy Sp. & 2.05 & 2.09 & 2.08 & 2.10 & 2.05 & 2.07 & 2.07 & 2.07 & 2.09 & 2.09 & 2.08 & 2.11 & 2.05 & 2.08\\ 
        % \noalign{\vskip 1mm}
        \cmidrule{2-17}
        & \multirow{3}{*}{\ding{56}} &DAS~\cite{delay_sum} & 2.07 & 2.10 & 2.10 & 2.11 & 2.06 & 2.07 & 2.07 & 2.08 & 2.09 & 2.10 & 2.09 & 2.13 & 2.06 & 2.09\\
        & &GSC~\cite{GSC}&2.12 & 2.15 & 2.14 & 2.14 & 2.08 & 2.08 & 2.07 & 2.09 & 2.12 & 2.13 & 2.13 & 2.17 & 2.11 & 2.12\\
        & &JNF~\cite{JNF}&2.07 & 2.11 & 2.10 & 2.11 & 2.06 & 2.08 & 2.08 & 2.09 & 2.10 & 2.10 & 2.09 & 2.13 & 2.07 & 2.09\\ 
        % \noalign{\vskip 1mm}
        \cmidrule{2-17}

        & \multirow{5}{*}{\ding{51}} &U-Net~\cite{unet}&2.41 & 2.44 & 2.42 & 2.47 & 2.47 & 2.44 & 2.12 & \textbf{2.42} & 2.48 & 2.49 & 2.48 & 2.52 & 2.46 & 2.43\\
        & &IPD U-Net&2.30 & 2.34 & 2.32 & 2.36 & 2.32 & 2.28 & 2.09 & 2.29 & 2.37 & 2.41 & 2.40 & 2.43 & 2.37 & 2.33\\
        & &BF U-Net&2.44 & 2.46 & 2.44 & 2.47 & 2.44 & 2.40 & 2.11 & 2.30 & 2.36 & 2.39 & 2.37 & 2.40 & 2.35 & 2.38\\
        & &CDUNet (fixed)&\textbf{2.53} & \textbf{2.57} & \textbf{2.56} & \textbf{2.57} & \textbf{2.57} & \textbf{2.42} & \textbf{2.13} & 2.40 & \textbf{2.59} & \textbf{2.55} & \textbf{2.56} & \textbf{2.59} & \textbf{2.52} & \textbf{2.50}\\
        & &CDUNet (varied)&2.39 & 2.44 & 2.45 & 2.47 & 2.44 & 2.37 & 2.12 & 2.35 & 2.46 & 2.46 & 2.44 & 2.48 & 2.42 & 2.41\\
        \midrule
        \midrule
        \multirow{9}{*}{5 dB}& - & Noisy Speech &2.34	&2.40	&2.38	&2.40	&2.35	&2.38	&2.38	&2.39	&2.38	&2.40	&2.38	&2.42	&2.34	&2.38 \\
        % \noalign{\vskip 1mm}
        \cmidrule{2-17}
        & \multirow{3}{*}{\ding{56}} &DAS~\cite{delay_sum}&2.36	&2.42	&2.40	&2.41	&2.37	&2.39	&2.38	&2.39	&2.39	&2.38	&2.42	&2.39	&2.44	&2.39 \\
        &&GSC~\cite{GSC}&2.41	&2.46	&2.45	&2.44	&2.39	&2.39	&2.38	&2.40	&2.41	&2.45	&2.44	&2.48	&2.41	&2.42 \\
        &&JNF~\cite{JNF}	&2.36	&2.42	&2.40	&2.42	&2.37	&2.40	&2.39	&2.40	&2.39	&2.42	&2.40	&2.44	&2.37	&2.40 \\
        % \noalign{\vskip 1mm}
        \cmidrule{2-17}
        &\multirow{5}{*}{\ding{51}}&U-Net~\cite{unet}&2.69&2.77&2.76&2.78&2.78&\textbf{2.75}&2.45&2.72&2.82&2.82&2.80&2.85&2.76&2.75\\
        &&IPD U-Net	&2.57	&2.63	&2.63	&2.64	&2.61	&2.60	&2.42	&2.60	&2.70	&2.73	&2.72	&2.76	&2.67	&2.64 \\
        &&BF U-Net &2.74&2.79&2.77&2.77&2.75&2.71&2.43&2.60&2.67&2.69&2.67&2.71&2.62&2.69\\
        &&CDUNet (fixed)&\textbf{2.89}&\textbf{2.94}	&\textbf{2.93}	&\textbf{2.92}	&\textbf{2.90}	&2.73	&\textbf{2.45}	&\textbf{2.76}	&\textbf{2.85}	&\textbf{2.85}	&\textbf{2.83}	&\textbf{2.86}	&\textbf{2.79}	&\textbf{2.82}\\
        &&CDUNet (varied)&2.72&2.79&2.78&2.78&2.75&2.68&2.45&2.66&2.78&2.77&2.76&2.81&2.72&2.73\\
        \bottomrule
    \end{tabular}
    \label{tab:main-table}
    \vspace{-1.0em}
\end{table*}
    \item \textbf{Mixture.} We synthesize the raw audio captured by the two microphones using the simulated RIR. For training, the ground-truth signal is obtained from the early-reverberated component of the pristine signal, with a reverb delay of 150 ms.
    % After determining the characteristics of the room, reverberation characteristics, and other details, we synthesize the raw audio captured by the two microphones using the simulated RIR. For training, the ground-truth signal is derived from the early-reverberated component of the pristine signal, with a reverb delay of 150 ms.
\end{itemize}
Based on this method, we synthesize two datasets for baseline models and the proposed CDUNet:
\begin{itemize}
    \item \textbf{Fixed-target dataset.} The fixed-target dataset is designed to compare our CDUNet model with the beamformer and other neural network baselines that don't accept angle as an input feature. This dataset contains 250,000 training samples, with the target direction fixed between 85° and 95°. The signal-to-noise ratio (SNR) of the target speech relative to the mixture of interfering speakers varies from -5 dB to 10 dB across the samples.
    % The target direction for these samples is confined between 85° and 95°, as depicted in figure \ref{simulator}. The interference source is positioned at an angle of 15° away from the target direction. 
    % The target direction for these samples is confined between 85° and 95°, as depicted in figure \ref{simulator}. The interference source is positioned at an angle of 15° away from the target direction.
    % The signal-to-noise ratio (SNR) of the target speech relative to the mixture of interfering speakers varies from -5 dB to 10 dB across the samples.
    \item \textbf{Variable-target dataset.}  
    To train a neural network for directed speech enhancement, we develop the variable-target dataset. This dataset introduces variability in the target speaker's location, which is randomly generated, while the interference direction remains 15° away from the target direction. The SNR and the number of utterances in this dataset are consistent with those in the fixed-target dataset.
    %For the purpose of training a neural network for directed speech enhancement, we develop a second dataset, designated as the variable-target dataset, which introduces variability in the target speaker's location.  The target direction for each sample is randomly generated, while the interference direction remains 15° away from the target direction. The SNR and the number of utterances in this dataset are consistent with those in the fixed-target dataset.
\end{itemize}

\vspace{-2pt}
\subsection{Model Setup}
\vspace{-1pt}
The input STFT and output inverse-STFT utilize a window size of 512 and a step size of 256.
The STFT representations are treated as separate channels and decomposed into their respective phase and magnitude components, culminating in a 10-channel frequency-domain input for the CDUNet model.

\vspace{-2pt}
\subsection{Evaluation}
\vspace{-1pt}
\textbf{Evaluation procedure.} We construct both the fixed-target and the variable-target evaluation datasets:
\begin{itemize}
    \item For the fixed-target evaluation, we create a corresponding test set with the target angle $\varphi_{target}=90^{\circ}$ and the interference angle span $\varphi_{inter}$ from $0^{\circ}$ to $180^{\circ}$, which consists of 500 utterances at SNR of 0 dB and 5 dB. 

\item For the evaluation of varied targets, we generate 500 utterances at 0 dB with the target direction $\varphi_{target}$ spanning from 0$^{\circ}$ to 90$^{\circ}$ and evaluate the PESQ scores. We apply such angle setup because 0°-90° is symmetrical to 90°-180°for the two-microphone model.

% For the evaluation of varied targets, we generate 500 utterances at 0 dB with the target direction spanning from 0$^{\circ}$ to 90$^{\circ}$ and evaluate the PESQ scores. We apply such angle setup because 0°-90° is symmetrical to 90°-180°for the two-microphone model.
\end{itemize}

\textbf{Baselines.} Two kinds of baselines are selected:
\begin{itemize}
    \item \textbf{U-Net free.} DAS is the traditional delay-and-sum beamformer model~\cite{delay_sum},  GSC is the generalized sidelobe canceller model~\cite{GSC} while JNF~\cite{JNF} stands for Joint Spatial and Tempo-Spectral Non-linear Filter, which was conducted using a circular array comprising three microphones, leading to superior results not captured in Table \ref{tab:main-table}.
    \item \textbf{U-Net based.} U-Net~\cite{unet} employs the same U-Net architecture as displayed in the figure\ref{overwiew}. The IPD Unet model incorporates the inter-microphone phase difference (IPD), a well-established spatial feature, as input to the U-Net architecture, and BF U-Net supplements the U-Net model with the output from the beamformer, which lacks the width input in comparison to our proposed model CDUNet.
\end{itemize}

\textbf{Evaluation metrics.} We evaluated the perceptual evaluation of speech quality (PESQ) for speech enhancement. Besides, the enhanced speech is fed into a pre-trained ASR model for word error rate (WER).

\section{Results and Analysis}

\subsection{Fixed Area and Directed Training}
\vspace{-3pt}
We evaluate the enhancement efficacy where the position of the target speaker remains constant while the direction of the interfering speech varies. All models are trained on our fixed-target dataset, which was designed with target speakers positioned at approximately 90°, aligning with the test conditions.
% The first row of Table \ref{tab:main-table} shows the perceptual evaluation of speech quality (PESQ) scores with the interference angles varying from 0 to 180 degrees. 
The first row of Table \ref{tab:main-table} of each SNR setup shows the PESQ scores with different interference angles. All of the U-Net based models outperform U-Net free ones, which reveals the superior efficiency of the U-Net architecture. Results on U-Net based models indicates that the inclusion of IPD significantly degrades performance, while the influence of the beamformer's output on performance is relatively negligible. The proposed CDUNet, with only 74.4k parameters, achieves markable improvement against baselines of U-Net structures when trained on the fixed-target dataset, which highlights the effectiveness of $\varphi_{width}$ in boosting the performance.
Moreover, the CUDNet trained on the variable-target dataset with changeable speaker locations still yields comparable results to the fixed-area enhancement as the U-Net. This achievement is particularly remarkable given its parameters of only 74.4K, while the JNF's contain 1M parameters. Therefore, our proposed CDUNet is capable of learning not only one spatial filter but 180 with much fewer (1400 instead of 250000) training examples per direction. This enhanced learning efficiency with fewer examples underscores the model's robustness and adaptability in processing complex spatial data.

% In conclusion, CDUNets achieved notably higher scores, outperforming baseline models while providing better orientation robustness. 
\vspace{-5pt}
\begin{table}[h]
    \centering
    \caption{PESQ scores for a fixed target of different input widths.}  
    \begin{resizebox}{1.0\columnwidth}{!}{
    \begin{tabular}{c|c|cccccc|c}
        \toprule
         \multirow{2}{*}{\textbf{Model}} & \multirow{2}{*}{$\bm{\varphi_{width}}$}  & \multicolumn{6}{c|}{$\bm{\varphi_{inter}}$} & \multirow{2}{*}{\textbf{Avg.}} \\
         \cmidrule{3-8}
           & & $0^{\circ}$ & $15^{\circ}$ & $30^{\circ}$ & $45^{\circ}$ & $60^{\circ}$ & $75^{\circ}$ & \\
        \midrule
        Noisy Sp. &-&2.05&2.09&2.08&2.10&2.05&2.07&2.07\\
        U-Net&-&2.41&2.44&2.42&2.47&2.47&2.44&2.44\\
        % \noalign{\vskip 0.5mm}
        \midrule
        \multirow{6}{*}{CDUNet}&3°&2.48&2.50&2.49&2.52&2.49&2.46&2.49\\
        &5°&2.49 & 2.53 & 2.52 & 2.52 & 2.51 & 2.44 & 2.50\\
        &7°&\textbf{2.53}&\textbf{2.57}&\textbf{2.56}&\textbf{2.57}&\textbf{2.57}&2.42&\textbf{2.54}\\
        &15°&2.50&2.52&2.51&2.54&2.51&\textbf{2.47}&2.51\\
        &20°&2.42 & 2.45 & 2.44 & 2.48 & 2.46 & 2.46&2.45\\
        &60°&2.34 & 2.37 & 2.37 & 2.41 & 2.41 & 2.40& 2.38\\    
        \bottomrule
    \end{tabular}
    }\end{resizebox}
    \label{tab:table-2}
    \vspace{-10pt}
\end{table}

    \vspace{-2pt}
\subsection{Different $\varphi_{width}$ for CDUNet}
To determine the optimal input width for the CDUNet, we evaluate PESQ scores with diverse input angles, utilizing the fixed-target training dataset.
We assess the impact of varying $\varphi_{width}$ on the enhancement of speech quality. Results in \ref{tab:table-2} indicate that optimal performance is attained with $\varphi_{width}=7^{\circ}$. The model employs beamforming based on the target direction and the edge angles. The edge angles are calculated by adjusting the target angle upwards and downwards by the specified input width. The inclusion of the width input enables the model to discern the spatial distribution of interfering speeches, which is achieved by comparing the edge speech signals with the target signals.

It's worth noticing that the angular separation between the target and interference direction always exceeds 15° within our dataset. Consequently, when the width is set below 15°, the model effectively separates the target speech by leveraging the input width as a discriminative boundary. On the contrary, when the width is narrowed to 3°, the proximity of the input width to the target direction leads to a diminished acquisition of novel and effective information, thus slightly damaging the model's performance. When the width extends beyond 15 degrees, the input width fails to separate the target direction from the interfering direction, resulting in degraded performance. In practical applications, the input width can be flexibly adjusted to align with the actual interference direction.

% To determine the optimal input width for the CDUNet, we evaluate PESQ scores with diverse setups, utilizing the identical training and testing datasets employed in the \hyl{"Fixed target speaker location"} experiment. % （这些内容是不是在前面的setup里面讲过了？如果讲过了这里可能需要删了)

% \begin{table}[h]
%     \centering
%     \caption{PESQ scores for a fixed training scheme of different widths. The target angle designated for the test is 90°, with the first line representing the interference angle.}  
%     \begin{resizebox}{1.0\columnwidth}{!}{
%     \begin{tabular}{c|cccccc|c}
%         \toprule
%         width & $0^{\circ}$ & $15^{\circ}$ & $30^{\circ}$ & $45^{\circ}$ & $60^{\circ}$ & $75^{\circ}$ & Avg.\\
%         \midrule
%         3&2.482&2.504&2.492&2.516&2.487&2.459&2.490\\
%         5&2.491 & 2.530 & 2.519 & 2.522 & 2.506 & 2.436 &\textbf{2.501}\\
%         7&\textbf{2.529}&\textbf{2.568}&\textbf{2.561}&\textbf{2.572}&\textbf{2.566}&2.418&\textbf{2.536}\\
%         % width10&2.442&2.473&2.449&2.477&2.455&2.381&2.446\\
%         15&2.495&2.523&2.513&2.537&2.509&\textbf{2.473}&\textbf{2.509}\\
%         20&2.422 & 2.450 & 2.444 & 2.483 & 2.462 & 2.455&2.453\\
%         % width45&2.582 & 2.613 & 2.615 & 2.611 & 2.597 & 2.422 & \textbf{2.573}\\
%         60&2.336 & 2.366 & 2.365 & 2.410 & 2.407 & 2.402& 2.381\\    
%         \bottomrule
%     \end{tabular}
%     }\end{resizebox}
%     \label{tab:table-2}
% \end{table}

\subsection{Directed Speech Enhancement}

% \begin{table}[h]
%     \centering
%     \caption{PESQ scores when the target speech source comes from different angles. The first line represents the target angle.}
%     % \begin{resizebox}{1.0\columnwidth}{!}{
%     \begin{tabular}{c|cccc|c}
%         \toprule
%         Model & $0^{\circ}$ & $30^{\circ}$ & $60^{\circ}$ & $90^{\circ}$& Avg.\\
%         \midrule
%         noise&2.105&2.071&2.040&2.089&2.076\\
%         \noalign{\vskip 1mm} % 在这里插入垂直间距
%         DAS&2.141&2.117&2.071&2.095&2.106\\
%         GSC&2.105&1.925&2.005&2.144&2.045\\
%         JNF&2.112&2.078&2.057&2.103&2.088\\
%         \noalign{\vskip 1mm} % 在这里插入垂直间距
%         U-Net&1.230&1.236&1.251&2.530&1.562\\
%         IPD\_U-Net&1.217&1.236&1.240&2.437&1.532\\
%         BF\_U-Net&2.054&1.960&1.943&2.272&2.057\\
%         CDUNet(ours)&\textbf{2.469}&\textbf{2.595}&\textbf{2.533}&\textbf{2.480}&\textbf{2.519}\\
   
%         \bottomrule
%     \end{tabular}
%     % }\end{resizebox}
%     \label{tab:table-3}
% \end{table}

\begin{table}[t]
    \centering
    \caption{PESQ scores with varied target speaker location.}
    \begin{resizebox}{0.85\columnwidth}{!}{
    \begin{tabular}{c|cccc|c}
        \toprule
        \multirow{2}{*}{\textbf{Model}} & \multicolumn{4}{c|}{\textbf{$\bm{\varphi_{target}}$}} & \multirow{2}{*}{\textbf{Avg.}} \\
        \cmidrule{2-5}
        & $0^{\circ}$ & $30^{\circ}$ & $60^{\circ}$ & $90^{\circ}$& \\
        \midrule
        Noisy Sp.&2.11&2.07&2.04&2.09&2.08\\
        % \noalign{\vskip 1mm} % 在这里插入垂直间距
        \midrule
        DAS&2.14&2.12&2.07&2.10&2.11\\
        GSC&2.11&1.93&2.01&2.14&2.05\\
        JNF&2.11&2.08&2.06&2.10&2.09\\
        % \noalign{\vskip 1mm} % 在这里插入垂直间距
        \midrule
        U-Net&1.23&1.24&1.25&2.53&1.56\\        
        IPD U-Net&1.22&1.24&1.24&2.44&1.53\\
        BF U-Net&2.05&1.96&1.94&2.27&2.06\\
CDUNet (ours)&\textbf{2.47}&\textbf{2.60}&\textbf{2.53}&\textbf{2.48}&\textbf{2.52}\\
       
        \bottomrule
    \end{tabular}
    }\end{resizebox}
    \vspace{-7pt}
    \label{tab:table-3}
\end{table}

Table \ref{tab:table-3} reports that our CDUNet model maintains its capacity to adapt dynamically to the target speaker, leveraging the input angle to achieve consistent enhancement effects across all directions, irrespective of the varying target angles. Although these baseline models are available to directional speech enhancement within a predefined target area, their performance deteriorates greatly when the target speaker's position shifts. This limitation necessitates the retraining of a model that is specifically calibrated to the new target direction, thereby incurring additional complexity and computational demands. Furthermore, this capability highlights the versatility of our model, as it can be effectively deployed across a multitude of real-world scenarios without the requirement for scenario-specific model training. This attribute exemplifies the model's robust adaptability and its potential for widespread application.

\subsection{Enhanced Signal for Downstream ASR}

We employ the same model and use the identical test dataset from the first Fixed target speaker location experiment to assess the model's performance. By evaluating the average performance across various test subsets, we aimed to quantify the model's enhancement effects on speech recognition tasks by using  the powerful open-sourced NeMo\footnote{\url{https://catalog.ngc.nvidia.com/orgs/nvidia/teams/nemo/models/stt_en_fastconformer_hybrid_large_pc}} ASR model. This experiment serves as a supplement to the fixed-target experiment, aiming at validating the enhancement of our model for downstream tasks. The results presented in Table \ref{tab:table-4} further underscore that
% despite varying degrees of distortion in the enhanced signals, such distortions did not exert an excessive negative impact on downstream tasks such as speech recognition. This further underscores that 
our proposed CDUNet model not only achieves high scores in speech quality but also exhibits excellent compatibility with downstream tasks.

\vspace{-9pt}

\begin{table}[h]
    \centering
    \caption{Performances of the downstream task for the fixed target speaker's location(training for target angle around $90^{\circ}$). } 
    \vspace{-5pt}
    \begin{resizebox}{1.0\columnwidth}{!}{
    \begin{tabular}{c|cc|cccccc}
        \toprule
        \multirow{2}{*}{\textbf{SNR}} & \multicolumn{8}{c}{\textbf{Wav Type (WER $\downarrow$)}} \\
        \cmidrule{2-9}
        & \textbf{Ref.} & \textbf{Noisy Sp.} & \textbf{DAS} & \textbf{GSC} & \textbf{JNF}& \textbf{IPD U-Net} & \textbf{U-Net} & \textbf{CDUNet (Ours)}\\
        \midrule
        \multirow{1}{*}{0 dB}
        %&PESQ&4.64&2.08&2.09&2.12&2.09&2.33&2.43&\textbf{2.50}\\
        % &STOI&1.00&0.86&0.86&\textbf{0.87}&0.77&0.80&0.81&0.81\\
        & 2.20 & 6.65 & 6.39 & 5.36 & 6.97 & 5.07 & 4.70 & \textbf{4.35} \\
        % \midrule
        \multirow{1}{*}{5 dB} 
        % &PESQ&4.64&2.38&2.39&2.42&2.40&2.64&2.75&\textbf{2.82}\\
        % &STOI&1.00&0.91&0.91&\textbf{0.92}&0.81&0.83&0.83&0.84\\
        & 2.20 & 3.93 & 3.84 & 3.46 & 4.14 & 3.46 & 3.37 & \textbf{3.11} \\
        \bottomrule
    \end{tabular}
    }\end{resizebox}
    
    \label{tab:table-4}
\end{table}

\vspace{-5pt}
\section{Conclusion}
In this research, we propose a directional multi-channel speech enhancement approach that integrates a beamformer with a causal U-Net model. The model processes raw microphone audio, using the output of  beamformer along with target and width parameters to achieve focused enhancement toward the target speaker. Our triple-steering spatial selection method not only effectively improves speech quality under multiple scenarios but also demonstrates superior performance for the downstream ASR task. Additionally, our model operates efficiently with just two microphones, making it suitable for real-world on-device applications.

% \begin{table}[h]
%     \centering
%     \vspace{-10pt}
%     \caption{An example of the result table. We can modify this to construct our own tables.}
%     \begin{resizebox}{1.0\columnwidth}{!}{
        
%     \begin{tabular}{c|cc|c|cccccc|c}
%         \toprule

%         \multirow{2}{*}{\textbf{ID}} & \multirow{2}{*}{{$\bm{\mathcal{L}_{ICTC}}$}} & \multirow{2}{*}{ \textbf{Decoding Alg.} }  & \multicolumn{7}{c|}{\textbf{SNR}} & \multirow{2}{*}{\textbf{Avg.}} \\
%         \cmidrule(lr){4-10}
%          & &  & -5 & 0 & 5 & 10 & 15 & 20 & +inf &  \\            
%         \midrule
%         % 4 FA
%         A & \multirow{2}{*}{\ding{56}} & KWS Graph & 41.6 & 75.1 & 90.2 & 95.8 & 97.5 & 98.3 & 98.8 & 85.3 \\ 
%         B &  & Streaming & 54.0 & 82.7 & 93.9 & 97.6 & 98.9 & 99.3 & 99.4 & 89.4 \\
%         \midrule
        
%        C &  \multirow{3}{*}{\ding{51}} & KWS Graph & 44.4 & 76.7 & 90.2 & 95.3 & 96.8 & 97.5 & 98.3 & 85.6 \\
%        D &  & Streaming & 59.8 & 86.4 & 95.4 & 98.2 & 98.9 & 99.2 & 99.7 & 91.1 \\
%                                    % & Multi-stage Streaming & \textbf{63.4} & \textbf{88.4} & \textbf{95.9} & \textbf{98.7} & \textbf{99.2} & \textbf{99.5} & \textbf{99.8} & \textbf{92.1} \\
%        % 注：这里填的是表3行(0, 30, 900)的结果
%        E & & M-S. Streaming  & \textbf{63.4} & \textbf{88.4} & \textbf{95.9} & \textbf{98.7} & \textbf{99.2} & \textbf{99.5} & \textbf{99.8} & \textbf{92.1} \\
%         \bottomrule
%     \end{tabular}
%     }\end{resizebox}

%     \label{tab:main-table}
% \end{table}

\bibliographystyle{style/IEEEtran}
\bibliography{refs/refs}

\end{document}